\documentclass[12pt,preprint]{aastex}

\slugcomment{submitted to ApJ,\today}

\shorttitle{Constraints on jet formation mechanisms}
\shortauthors{Li \& Cao}

\begin{document}

\title{Constraints on jet formation mechanisms with the most energetic giant outbursts in MS 0735+7421}

\author{Shuang-Liang Li and Xinwu Cao}

\altaffiltext{}{Key Laboratory for Research in Galaxies and
Cosmology, Shanghai Astronomical Observatory, Chinese Academy of
Sciences, 80 Nandan Road, Shanghai, 200030, China; lisl@shao.ac.cn,
cxw@shao.ac.cn}

\begin{abstract}

Giant X-ray cavities lie in some active galactic nuclei (AGNs)
locating in central galaxies of clusters, which are estimated to
have stored $10^{55}\sim 10^{62}$~erg of energy. Most of these
cavities are thought to be inflated by jets of AGNs on a timescale
of $\ga 10^7$ years. The jets can be either powered by rotating
black holes or the accretion disks surrounding black holes, or both.
The observations of giant X-ray cavities can therefore be used to
constrain jet formation mechanisms. In this work, we choose the most
energetic cavity, MS 0735+7421, with stored energy $\sim
10^{62}$~erg, to constrain the jet formation mechanisms and the
evolution of the central massive black hole in this source. The
bolometric luminosity of the AGN in this cavity is $\sim 10^{-5}
L_{\rm Edd}$, however, the mean power of the jet required to inflate
the cavity is estimated as $\sim 0.02L_{\rm Edd}$, which implies
that the source has experienced strong outbursts previously. During
outbursts, the jet power and the mass accretion rate should be
significantly higher than its present values. We construct an
accretion disk model, in which the angular momentum and energy
carried away by jets is properly included, to calculate the spin and
mass evolution of the massive black hole. In our calculations,
different jet formation mechanisms are employed, and we find that
the jets generated with the Blandford-Znajek (BZ) mechanism are
unable to produce the giant cavity with $\sim 10^{62}$~erg in this
source. Only the jets accelerated with the combination of the
Blandford-Payne (BP) and BZ mechanisms can successfully inflate such
a giant cavity, if the magnetic pressure is close to equipartition
with the total (radiation$+$gas) pressure of the accretion disk. For
dynamo generated magnetic field in the disk, such an energetic giant
cavity can be inflated by the magnetically driven jets only if the
initial black hole spin parameter $a_0\ga0.95$. Our calculations
show that the final spin parameter $a$ of the black hole is always
$\sim 0.9-0.998$ for all the computational examples which can
provide sufficient energy for the cavity of MS 0735+7421.
\end{abstract}


\section{INTRODUCTION}

Giant X-ray cavities have been discovered in some central galaxies
of clusters by XMM-Newton and Chandra [e.g., Hydra A \citep{m2000};
RBS 797 \citep{s2001} and MS 0735+7421 \citep{m2005}], which contain
huge energy up to $10^{55} \sim 10^{62} $~erg. They can effectively
suppress the cooling of the intracluster medium (ICM) and provides a
direct evidence for the presence of AGN feedback in galaxy formation
and evolution \citep*[e.g.][]{b2004,m2005,a2006,c2006}. It is
believed that the cavities are inflated by the jets launched from
the accretion disks in active galactic nucleus (AGNs)
\citep*[see][for a review and the referneces
therein]{2007ARA&A..45..117M}. Therefore, the X-ray cavities provide
a direct measurement of the mechanical energy released by the jets
through the work done on the hot gas surrounding them. Measurements
of this energy, combined with measurements of the timescale required
to inflate the cavities, can be used to estimate the mean jet power
\citep*[e.g.,][]{a2006}.

The jet formation mechanisms in AGNs have been extensively studied
in the last several decades. The currently most favored jet
formation mechanisms contain two categories, i.e., the
Blandford-Znajek (BZ) process \citep{b1977} and the Blandford-Payne
(BP) process \citep{b1982}. In the BZ process, energy and angular
momentum are extracted from a rotating black hole and transferred to
a remote astrophysical load by open magnetic field lines. In the BP
process, the magnetic field threading the disk extracts energy from
the rotating gas in the accretion disk to power the jet/outflow. The
hybrid model proposed by \citet{m1999}, as a variant of the BZ
model, which combines the BZ and BP effects through the large scale
magnetic field threading the accretion disk outside the ergosphere
and the rotating plasma within the ergosphere.
The ordered large-scale magnetic field threading the disk is a
crucial ingredient in jet formation models. In most previous works,
the strength of the magnetic field is assumed to scale with the
gas/radiation pressure of the accretion disk
\citep*[e.g.,][]{m1996,g1997,l1999,1999ApJ...523L...7A,n2007,2008ApJ...687..156W,m2009,w2011}.
\citet{l1999} pointed out that even the calculations of
\citet{g1997} have overestimated the power of the Blandford-Znajek
process, since they have overestimated the strength of the
large-scale field threading the inner region of an accretion disk,
and then the power of the Blandford-Znajek process. The strength of
the large-scale field scales with the disk thickness, and it is very
weak if the field is created by dynamo processes for thin disk cases
\citep{1996MNRAS.281..219T,l1999}. The maximal jet power extracted
from an accretion disk can be estimated on the assumption that the
toroidal field component is of the same order of the poloidal field
component at the disk surface \citep*[see,
e.g.,][]{1992MNRAS.259P..23L,1993noma.book.....B}.


The giant cavity in MS 0735+7421, one of the most energetic
outbursts in AGNs, which stores a total $\sim 10^{62}$~erg of energy
with the timescale required to inflate the cavity $\sim 10^8$~years
\citep{m2005,m2009}. It implies that the mean jet power should be
$\sim 10^{46}$~erg ${\rm s} ^{-1}$ \citep{m2005,g2007}, which is
about two percent of the Eddington luminosity ($L_{\rm Edd}$) of a
$5 \times 10^9 M_{\odot}$ ($M_{\odot}$ is the solar mass) black hole
in MS 0735+7421 \citep{m2009}. However, the central AGN in this
cavity shows very low optical nuclear emission ($L_{\rm I} < 2.5
\times 10 ^{42}$~erg ${\rm s}^{-1} (L_{\rm I}$ is the luminosity in
I band), which implies its accretion disk accreting at the current
rate being too low to power strong jets with $\sim 10^{46}$~erg
${\rm s} ^{-1}$. \citet{m2009,m2011} argued that the BP process
solely is insufficient to power the giant outbursts in this source.
They suggested that the jet is alternatively powered by the
rotational energy of a extremely Kerr black hole ($\sim 10^{62}$
erg), which roughly corresponds to the total energy stored in this
cavity. However, \citet{l1999} pointed out that the magnetic field
strength near the horizon of a rotating black hole is dominantly
determined by the structure of the inner region of the disk. Thus,
it is still quite doubtful if such a faint accretion disk can
maintain a strong field near the horizon to power powerful jets with
$\sim 10^{46}$~erg ${\rm s} ^{-1}$ .



It is well known that quasars are powered by accretion onto massive
black holes, and the growth of massive black holes could be
dominantly governed by mass accretion in quasars. The massive black
holes are therefore the AGN relics \citep{1982MNRAS.200..115S}. The
faint AGN in MS 0735+7421 contains a massive black hole with mass of
$5 \times 10^9 M_{\odot}$, which may probably have experienced a
quasar phase before. During its quasar phase, the black hole was
accreting at relatively high rates, and therefore powerful jets can
be magnetically accelerated from the region near the black hole
either by the BP or BZ precesses, or both. In the same period of
time, the central massive black hole grew up, and the hole was spun
up simultaneously. In this work, we choose the most powerful cavity,
MS 0735+7421, to constrain the jet formation mechanisms and the
mass/spin evolution of the massive black hole in this source.


\section{MODEL}\label{models}

\subsection{The equations of accretion disk}

The faint AGN in MS 0735+7421 may probably have experienced a quasar
phase before. During its quasar phase, the black hole was accreting
at relatively high rates, and therefore its accretion disk can be
described by the standard thin disk model \citep{s1973}. In this
work, we consider a relativistic thin accretion disk around a Kerr
black hole \citep{n1973,m2000b}, and the metric around the black
hole reads (the geometrical unit $G=c=1$ is adopted):
\begin{equation}
ds^{2}=-\frac{r^2 \Delta}{A} dt^2+\frac{A}{r^2}(d\phi-\omega
dt)^2+\frac{r^2}{\Delta}dr^2+dz^2, \label{metric}
\end{equation}
\begin{displaymath}
\Delta=r^2-2Mr+a^2,
\end{displaymath}
\begin{displaymath}
A=r^4+r^2 a^2+2Mra^2,
\end{displaymath}
\begin{displaymath}
\omega=\frac{2Mar}{A},
\end{displaymath}
\begin{displaymath}
a=\frac{J}{M},
\end{displaymath}
where $M$ is the mass of the black hole, $J$ and $a$ are the angular
momentum and specific angular momentum of the black hole
respectively, and $\omega$ is the dragging angular velocity of the
metric.


The model of the geometrically thin, optically thick and Keplerian
accretion disk surrounding a Kerr black hole was developed by
\citet{n1973}. The structure of the accretion disk may be altered in
the presence of the large-scale magnetic field threading the disk,
which can drive the outflows/jets from the disk, and a fraction of
the angular momentum and energy of the disk is carried away in the
outflows/jets \citep*[e.g.,][]{1998ApJ...499..329O,c2002}. In order
to explore the magnetically accelerated jets from the accretion
diks, we properly consider the effects of the jets on the accretion
disk. The basic equations of a relativistic thin disk with
magnetically driven outflows/jets are summarized as follows.

The continuity equation is
\begin{equation}
\dot{M}=-2\pi \Delta^{1/2} \Sigma v_{\rm r}, \label{continuity}
\end{equation}
where $v_r$ is the radial velocity of the accretion flow,
$\Sigma=2\rho H$ is the surface density , and $\dot{M}$ is the mass
accretion rate. As we consider fast moving outflows/jets in this
work, the mass loss rate in the ouflow/jet is neglected in the
continuity equation of the accretion flow.

The radial momentum equation is
\begin{equation}
\frac{\gamma_\phi A M}{r^4 \Delta}\frac{(\Omega-\Omega_{\rm k}
^+)(\Omega-\Omega_{\rm k} ^-)}{\Omega_{\rm k} ^+ \Omega_{\rm k}
^-}+g_{\rm m}=0, \label{momentum}
\end{equation}
where $\Omega$ is the angular velocity, and the Lorentz factor
$\gamma_\phi$ of the rotational velocity $v_\phi$ is given by
\begin{displaymath}
\gamma_\phi=(1-v_\phi^2)^{-1/2},
\end{displaymath}
\begin{displaymath}
v_\phi=A\tilde{\Omega}/r^2 \Delta^{1/2},
\end{displaymath}
and $\tilde{\Omega}=\Omega-\omega$. In this work, we consider the
geometrically thin disks, and the gas pressure gradient in the
radial direction can be omitted in the radial momentum equation of
the disk \citep{s1973}. The first term in Equation (\ref{momentum})
represents the net force exerted on the gas moving with a circular
angular velocity $\Omega$ in the Kerr metric.  It should be balanced
with the radial magnetic force $g_{\rm m}$ induced by the
large-scale magnetic field threading the accretion disk without
considering the radial pressure gradient in the thin disk case. The
Keplerian angular velocities of the prograde ($+$) and retrograde
($-$) motions are
\begin{displaymath}
\Omega_{\rm k}^{\pm}=\pm\frac{M^{1/2}}{r^{3/2}\pm a M^{1/2}},
\end{displaymath}
and the radial magnetic force is
\begin{equation}
g_{\rm m}={B_{\rm r}B_{\rm z}}/{2\pi\Sigma},
\end{equation}
where  $B_{r}$ and $B_{z}$ are the radial and vertical components of
the magnetic fields at the disk surface, respectively. The
inclination of the field line at the disk surface $\kappa_0$ is
defined as $\kappa_0=B_z/B_r$.

The angular momentum equation is
\begin{equation}
-\frac{\dot{M}}{2\pi} \frac{dL}{dr} + \frac{d}{dr}(r
W^r_\phi)+T_{\rm m}r=0, \label{angular}
\end{equation}
where the angular momentum of the accretion flow $L$ is
\begin{displaymath}
L=\frac{A^{1/2}(\gamma^{2}_{\phi}-1)^{1/2}}{r},
\end{displaymath}
and the height-integrated viscous tensor is
\begin{displaymath}
W^r_\phi=\alpha \frac{A^{3/2}\Delta^{1/2}\gamma _{\phi}^3}{r^6} W,
\end{displaymath}
where $\alpha$ is the Shakura-Sunyaev viscosity parameter. The
height-integrated pressure $W=2HP_{\rm tot}$, where the total
pressure $P_{\rm tot}=P_{\rm gas}+P_{\rm rad}+P_{\rm m}$, $P_{\rm
gas},P_{\rm rad}$ and $P_{\rm m}=(1-\beta)P_{\rm tot}$ are the gas
pressure, radiation pressure and magnetic pressure respectively
($1-\beta$ is the ratio of the magnetic pressure to the total
pressure). The scale height $H$ of the accretion disk is given by
\begin{equation}
H^2=c_{s}^2 r^4/(L^2-a^2),
\end{equation}
where $c_{\rm s}=\sqrt{P_{\rm tot}/\rho}$ is the sound speed of the
gas in the disk.

The first two terms in Equation (\ref{angular}) represent the change
rate of angular momentum and the transfer rate of angular momentum
caused by the viscous tensor in the gas, and the third term is the
rate of angular momentum carried away by the magnetically driven
outflows/jets. The magnetic torque exerted on the accretion flow due
to the outflows/jets is
\begin{equation}
 T_{\rm m}=-{\frac {B_{\rm z} B_{\rm \varphi} R}{2\pi}}, \label{tm}
\end{equation}
where $B_{\rm \varphi}$ is the toroidal component of the field
strength at the disk surface. Equation (\ref{tm}) is the vertical
integration of magnetic torque in the ideal MHD angular momentum
equation \citep{c2002}.


The energy equation is
\begin{equation}
\nu \Sigma \frac{\gamma_{\phi}^4 A^2}{r^6}\left(
\frac{d\Omega}{dr}\right)^2= \frac{16acT^4}{3\bar{\kappa}\Sigma},
\label{energy}
\end{equation}
where $\nu$ is the viscosity coefficient, $\nu \Sigma
d\Omega/dr=-\alpha W/r$ in $\alpha$-viscosity,  and $T$ is the
temperature of the gas in the disk \citep{1996ApJ...471..762A}. The
left term in this equation is the surface heat generation rate
caused by turbulence of the gas in the disk, which is balanced with
the cooling rate of the disk. The radial advection of energy is
neglected in the thin disk case. The opacity $\bar{\kappa}$ of the
gas is given by
\begin{displaymath}
\bar{\rm \kappa}=\kappa_{\rm es}+\kappa_{\rm ff}=0.4+0.64\times
10^{23}\rho T^{-7/2} \textmd{cm}^2 \textmd{g}^{-1}, \label{opacity}
\end{displaymath}
where $\kappa_{\rm es}$ and $\kappa_{\rm ff}$ are the electron
scattering opacity and free-free opacity respectively.

\subsection{The jet formation mechanisms}




In the general form of the BZ mechanism, the jet power $L_{\rm BZ}$
can be estimated with \citep{g1997}
\begin{equation}
L_{\rm BZ}=\frac{1}{32}\omega^{2}_{\rm F}B^2_\perp r^2_{\rm H}
(J/J_{\rm max})^2 c, \label{lbz}
\end{equation}
where $r_{\rm H}$ is the horizon radius, $B_\perp$ is the component
of the magnetic field normal to the black hole horizon, $J$ and
$J_{\rm max}=GM^2/c$ are the angular momentum and maximum angular
momentum of a black hole, and $\omega_{\rm F}^2\equiv \Omega_{\rm F}
(\Omega_{\rm H}-\Omega_{\rm F})/\Omega_{\rm H}^2$ is a factor at the
black hole horizon determined by the angular velocity of black hole
and that of the magnetic filed lines. In this work, we simply adopt
$B_\perp \sim B$ ($B^2=B_{p}^2+B_{\varphi}^2$), which is the maximal
magnetic field strength in the disk \citep*[see][for the
details]{g1997}. It is easy to conclude that the maximal jet power
$L_{\rm BZ}^{\rm max}$ corresponds to $\Omega_{\rm F}=1/2
\Omega_{\rm H}$.
The power of the jets accelerated from an accretion disk can be
calculated with \citep*[e.g.,][]{l1999,c2002b},
\begin{equation}
L_{\rm BP}=\int_{r_{\rm in}}^{r_{\rm out}} \frac{B_{\rm p}B_{\rm
\varphi}}{4\pi} r \Omega 2\pi r {\rm d} r, \label{lbp}
\end{equation}
where $B_{\rm p}$ \textbf{($B_{p}^2=B_{r}^2+B_{z}^2$)} is the
poloidal field strength, and $r_{\rm in}$ and $r_{\rm out}$ are the
inner and outer radius of the accretion disk respectively.

\citet{1996MNRAS.281..219T} suggested that the typical size of the
magnetic fields produced by dynamo processes is roughly around the
disk thickness $H$. The large-scale field can be produced from the
small-scale field created by dynamo processes as $B(\lambda)\propto
\lambda^{-1}$ for the idealized case, where $\lambda$ is the
length-scale of the field. For the magnetically launching problem
for jets, the size of magnetic field lines $\ga R$ is required in
order to accelerate the jets/outflows efficiently. Thus,
\citet{l1999} proposed that the poloidal magnetic fields $B^{'}$
should be $\sim (H/r) B$ if they are generated through dynamo
processes in the accretion disk.  The jet power of the BZ and BP
mechanisms is given by
\begin{equation}
L_{\rm BZ}=\frac{1}{32}\omega^{2}_{\rm F} \left(B_\perp
\frac{H}{r}\right)^2_{\rm max} r^2_{\rm H} (J/J_{\rm max})^2 c,
\label{lbzlivio}
\end{equation}
and
\begin{equation}
L_{\rm BP}=\int_{r_{\rm in}}^{r_{\rm out}} \frac{B_{\rm p}B_{\rm
\varphi}}{4\pi} r \Omega 2\pi r \left(\frac{H}{r}\right)^2 {\rm d}
r, \label{lbplivio}
\end{equation}
on the assumption of magnetic field generated through dynamo
processes, where $(B_\perp {H}/{r})_{\rm max}$ is the maximal
$(B_\perp H/r)$ in the inner region of the accretion disk.

The difficulties in estimating the jet power, either for the BP or
the BZ processes, are the magnetic field strength and the field
geometry, which are still highly uncertain. The conventional
estimates of the field strength are more or less based on the
assumption of equipartition between magnetic pressure and gas
pressure/radiation pressure
\citep*[e.g.,][]{m1996,g1997,l1999,1999ApJ...523L...7A,n2007,2008ApJ...687..156W,m2009,w2011}.
In this work, we adopt a parameter $\beta$ to describe the field
strength in the disk. The magnetic pressure $P_{\rm
m}=(1-\beta)P_{\rm tot}$ ($P_{\rm tot}=P_{\rm gas}+P_{\rm rad}$),
and therefore the field strength $B=\sqrt{(1-\beta)8\pi P_{\rm
tot}}$. In principle, the magnetic field strength and the structure
of the accretion disk are available by solving a set of dynamical
equations of the disk described in Section 2.1 simultaneously, if
the value of $\beta$ and the field geometry are known. However, the
geometry of the magnetic field of the disk is in principle a global
problem \citep*[see, e.g.,][for the detailed discussion, and the
references therein]{2011AIPC.1381..227S}. It depends not only on the
initial strength and geometry of the magnetic field as indicated by
some MHD simulations \citep*[e.g.,][]{i2003,b2008,mb2012}, but also
on the processes of diffusion and advection in accretion disks
\citep{l1994,c2011,mb2012}, which is beyond the scope of this paper.
The inclination of the field line at the disk surface plays a key
role in launching jets. More specifically, the angle of field lines
inclined to the mid-plane of the disk is required to be less than
$60^{\circ}$ for launching jets from a Keplerian cold disk
\citep{b1982}. This critical angle can be slightly larger than
$60^{\circ}$ if the internal energy of the gas in the disk is
considered \citep{1994A&A...287...80C}. This critical angle could be
larger than $60^\circ$ even for the cold gas magnetically launched
from the inner edge of the accretion disk very close to a rapidly
spinning black hole \citep{1997MNRAS.291..145C}, which indicates
that the spin of the black hole may help to launch jets
centrifugally by cold magnetized disks
\citep{1997MNRAS.291..145C,2010A&A...517A..18S}. This is because the
frame drag effect of the spinning black hole may help to accelerate
the outflows. For realistic cases, the internal energy of the gas is
included in the Bernoulli equation of the outflow, and an additional
force due to the pressure gradient helps to accelerate the gas in
the outflow \citep*[e.g., see equation 14
in][]{1994A&A...287...80C}. In this case, the outflow can be
launched if the inclination angle is slightly larger than the
critical value. As the final derived accretion disk structure is
insensitive to the precise value of the field inclination at the
disk surface, we simply adopt the inclination of $60^\circ$ in all
our calculations. The strength of the azimuthal component of the
field at the disk surface is mainly determined by the properties of
the magnetically jets. The model properly including both the
accretion disk solution and jet solution is very complicated, which
is beyond the scope of this work. It was pointed out that the fast
moving jets/outflows always correspond to the case $B_\varphi\ll
B_p$, which $B_\varphi\sim B_p$ is satisfied for slowly moving
outflows with relatively high mass loss rate
\citep{1994A&A...287...80C,1998ApJ...499..329O,c2002,c2002b}. In the
case of the problem considered in this work, we focus on the fast
moving jets from the disk, and $B_\varphi\ll B_p$ should be
satisfied. In this work, we adopt $\xi_\varphi=0.1$
($B_\varphi=\xi_\varphi B_p$) in most of our calculations.



In summary, we calculate the jet power $L_{\rm jet}$ with two models
in this work, i.e.,
\\

Model A (general BZ$+$BP mechanisms): $L_{\rm jet}=L_{\rm BZ}+L_{\rm
BP}$, where $L_{\rm BZ}$ and $L_{\rm BP}$ are calculated with
Equations (\ref{lbz}) and (\ref{lbp}) respectively;
\\

Model B (Livio's model): $L_{\rm jet}=L_{\rm BZ}+L_{\rm BP}$, where
$L_{\rm BZ}$ and $L_{\rm BP}$ are calculated with Equations
(\ref{lbzlivio}) and (\ref{lbplivio}) respectively.
\\

The main difference between model A and B is the estimate of the
field strength in the accretion disk. Finally, the energy released
through BZ$+$BP mechanism during the outbursts is available with
\begin{equation}
E_{\rm tot}=E_{\rm BZ}+E_{\rm BP}=\int_0^{t} L_{\rm jet} {\rm d} t,
\label{etot}
\end{equation}
where $t$ is the duration of the AGN outbursts.

\subsection{The evolution of the black hole}


For a massive black hole surrounded by an accretion disk, the black
hole mass and spin evolution is described by
\citep*[e.g.,][]{m1996,1996ApJ...472..564L}
\begin{equation}
\frac{\textmd{d}a}{\textmd{d}t}=\frac{\dot{M}}{M}(\tilde{j}_{\rm
ms}-2a\tilde{e}_{\rm ms})-\frac{L_{\rm
BZ}}{Mc^2}\left(\frac{1}{k\tilde{\Omega}_{\rm h}}-2a\right),
\label{spin}
\end{equation}

\begin{equation}
\frac{\textmd{d} \ln M}{\textmd{d}t}=\frac{\dot{M}}{M}\tilde{e}_{\rm
ms}-\frac{L_{\rm BZ}}{Mc^2}, \label{mass}
\end{equation}
where a ($0\le a<1$) is the dimensionless angular momentum of black
hole, $\dot{M}\equiv \textmd{d}M/\textmd{d}t$ is the accretion rate,
$M$ is the mass of black hole, $\tilde{j}_{\rm ms}$ and
$\tilde{e}_{\rm ms}$ are the dimensionless specific angular momentum
and energy of the gas in the accretion disk at the marginally stable
orbit respectively, $\tilde{\Omega}_{\rm h}$ is the dimensionless
angular velocity at the horizon of the black hole, $L_{\rm BZ}$ is
the jet power due to the BZ process, and $k\equiv \Omega_{\rm
F}/\Omega_{\rm h} < 1$ is a constant, where $\Omega_{\rm F}$ is the
angular velocity of magnetic field lines at the horizon.

The time evolution of mass accretion rate is still quite unclear,
which may be dependent on the circum-nuclear gas near the black hole
or/and feedback of the quasar at the center of the galaxy
\citep*[e.g.,][]{2005Natur.433..604D}. The observed Eddington ratio
distribution for AGNs can be roughly described by a power-law
distribution, which is consistent with the self-regulated black hole
growth model. In this model, the feedback of AGNs produces a
self-regulating ¡°decay¡± or ¡°blowout¡± phase after the AGN reaches
some peak luminosity and begins to expel gas and shut down accretion
\citep{2005ApJ...630..716H,2005ApJ...625L..71H,h2009}. \citet{h2009}
suggested the quasar luminosity (then the accretion rate) evolving
with time can be described by
\begin{equation}
\frac{{\rm d} t}{{\rm dlog}\dot{m}}=-\tau_{\rm
Q}\left(\frac{\dot{m}}{\dot{m}_{\rm 0}} \right)^{{-\beta}_{\rm L}}
{\rm exp}\left(-\frac{\dot{m}}{\dot{m}_{\rm 0}} \right),
\label{ratio}
\end{equation}
where $\dot{m}\equiv \dot{M}/\dot{M}_{\rm Edd}$ is the accretion
rate in units of Eddington accretion rate ($\dot{M}_{\rm
Edd}=1.5\times 10^{18} M/M_{\odot} ~{\rm g}~ {\rm s^{-1}}$ is the
Eddington accretion rate), $\tau_{\rm Q}$ is a constant and
$\dot{m}_{\rm 0}$ is the peak accretion rate of quasar. The index
$\beta_{L} \sim 0.6$ is adopted as suggested by \citet{h2009}. We
define the lifetime $t_{\rm Q}$ of quasar as
\begin{equation}
t_{\rm Q}=\int_{0}^{t_{\rm Q}}dt=-\int_{\dot{m}_{\rm
0}}^{0.01}\tau_{\rm Q}\left(\frac{\dot{m}}{\dot{m}_{\rm 0}}
\right)^{{-\beta}_{\rm L}} {\rm
exp}\left(-\frac{\dot{m}}{\dot{m}_{\rm 0}} \right) {\rm
dlog}\dot{m}, \label{tq}
\end{equation}
where the integral upper limit $\dot{m}=0.01$ is adopted as the
typical lower limit on broad-line AGNs
\citep*[e.g.,][]{2006ApJ...648..128K,2011ApJ...733...60T}. We assume
$\dot{m}_0=1$ in all our calculations. The quasar lifetime $t_{\rm
Q}\sim 10 \tau_{\rm Q}$ for $\beta_{L}=0.6$.

\section{RESULTS}\label{results}

As described in Sections 2.1 and 2.2, the structure of an accretion
disk surrounding a rotating black hole with mass $M$ and spin
parameter $a$ can be calculated by solving Equations
(\ref{continuity})$-$(\ref{energy}) for different jet formation
mechanisms, provided the values of the disk parameters, $\alpha$,
$\beta$, and $\dot{M}$, are supplied. Based on the derived disk
structure and jet power, the mass and spin evolution of the black
hole can be calculate with Equations (\ref{spin}) and (\ref{mass}),
if suitable initial conditions and the mass accretion rate as a
function of time are specified. {The conventional value of viscosity
parameter $\alpha=0.1$ is adopted in all our calculations. In Figure
\ref{property}, we find that the disk structure has been
significantly altered by the magnetically accelerated jets. The
temperature of the accretion disk with outflows/jets is obviously
lower than that of the standard thin accretion disk, because a
fraction of the gravitational energy released in the disk is carried
away in the outflows/jets. The temperature of accretion disk
decreases significantly for model A than that for model B, because
the jet power is higher for model A if all other model parameters
are fixed (see Figure \ref{property}).



The central AGN in the cavity of MS 0735+7421 contains a $5 \times
10^9 M_{\odot}$ black hole, which shows very low optical nuclear
emission ($L_{\rm I} < 2.5 \times 10 ^{42}$~erg ${\rm s}^{-1}$ in I
band). The model calculations for this source should satisfy three
observational quantities, total energy stored in the cavity $\sim
10^{62}$~erg, present mass of the central black hole $5 \times 10^9
M_{\odot}$, and the timescale of the jet activity in this AGN
$\sim10^8$~years \citep{m2005,m2009}. It implies that the giant
cavity in this source was inflated by the jets when the black hole
was accreting at relatively high rates previously. In all our
calculations, the initial black hole mass $M_0$ is chosen in such a
way to let the final black hole mass be $5 \times 10^9 M_{\odot}$ as
that in the AGN of MS 0735+7421. Thus, the time-dependent mass
accretion rate $\dot{m}(t)$ is a crucial ingredient in our model
calculations. In this work, we adopt a power-law time dependent
$\dot{m}(t)$ with an exponential cutoff at the high end $\dot{m}_0$
suggested by \citet{h2009}. It is found that the mass accretion rate
declines very rapidly for this time-dependent $\dot{m}(t)$ when
$\tau_{\rm Q}$ is small, i.e., a short quasar lifetime (see Figure
\ref{edd}).

The total energy released during the outbursts as functions of time
with different values of the model parameters for different jet
formation models are plotted in Figures \ref{compare}-\ref{beta}. In
Figures \ref{compare} and \ref{compareb}, we plot the energy
released during the outbursts as functions of the outburst time with
different sets of the model parameters for different initial black
hole spin parameters $a_0=0.1$ and $0.95$, respectively. It is found
that the total energy released in the cavity is higher for model A
than that for model B. The total energy increases with $\tau_{\rm
Q}$, and much energy is carried away in the jets provided the
initial black hole spin parameter is high, if all other parameters
are fixed.

The black hole mass and bolometric luminosity of broad-line AGNs can
be well estimated with the observations of the SEDs and broad-line
width, and therefore the Eddington ratio distribution of broad-line
AGNs is available. The peaks of the Eddington ratio distributions
derived with different broad-line AGN samples are around $0.1-0.4$
\citep*[e.g.,][]{2004MNRAS.352.1390M,2004ApJ...608..136W,2006ApJ...648..128K,2009MNRAS.397..135K,s2012}.
The time evolution of accretion rate is still quite uncertain. A
constant accretion rate is usually assumed for AGNs in most of the
previous studies of the cosmological evolution of massive black
holes
\citep*[e.g.,][]{2002MNRAS.335..965Y,2004MNRAS.351..169M,2004MNRAS.354.1020S}.
The jet power is dependent on the mass accretion rate $\dot{m}$, and
for comparison, we alternatively adopt a constant mass accretion
rate, $\dot{m}=0.25$, which is around the peak value of the
distribution in the sample of \citet{2006ApJ...648..128K}, to
calculate the total energy released in the cavity as functions of
time (see Figure \ref{compare2}).

We compare the relative importance of the BP and BZ mechanisms in
Figure \ref{bzbp}. Similar to that suggested by \citet{l1999}, we
find that the BP power always dominates over the BZ power even if
the black hole is rotating rapidly. The results calculated with
different values of $\beta$ are plotted in Figure \ref{beta}. The
jet power decreases significantly with increasing $\beta$. In most
of our calculations, $\beta=0.5$ is adopted, which is almost the
upper limit on the magnetic field strength, and then the jet power.

The evolution of black hole spin for different jet formation models
is given in Figure \ref{spin2}.


\section{DISCUSSION}\label{sec:summary}

The physics of jet formation is still quite unclear. As described in
the previous sections, the total energy released by the jets of the
AGN and the mass/spin evolution of the central massive black hole
can be calculated with suitable initial conditions for a given jet
formation model. The most powerful outbursts observed in MS
0735+7421 provide constraints on different jet formation models.

The basic physics of the jet formation models adopted in this work
is the same as the previous works of \citet{g1997} and
\citet{l1999}, however, their estimates of the  magnetic field
strength of the disk are based on the conventional accretion disk
models without magnetic field. This is a good approximation in the
weak magnetic field case, as the disk structure has not been altered
significantly by the field, while the assumption becomes invalid for
strong field cases. In this work, we calculate the structure of the
accretion disk including the magnetic torque exerted on the disk and
the energy carried away by the jets (see Section 2.1).

The giant cavity in MS 0735+7421 stores a total $\sim 10^{62}$~erg
of energy, which is the most energetic outbursts of AGN so far
discovered, if the cavity is inflated by the jets of the AGN. The
observations of MS 0735+7421 provide very useful information that
can constrain the detailed jet formation models. A power-law time
dependent $\dot{m}(t)$ with an exponential cutoff at the high end is
adopted in most of our calculations, which is derived based on the
scenario of the feedback of AGNs producing a self-regulating
¡°decay¡± or ¡°blowout¡± phase after the AGN reaches some peak
luminosity \citep{2005ApJ...630..716H,2005ApJ...625L..71H,h2009}. It
is found that the total energy released in the jets can be as high
as $\ga 10^{62}$~erg only in the calculations with model A, i.e, the
strength of large-scale magnetic field fields scales directly with
the total pressure of the accretion disk, if the initial black hole
spin parameter $a_0=0.1$ is adopted (see Figure \ref{compare}). The
another jet formation model seems not to be able to produce such a
giant cavity containing $\sim 10^{62}$~erg even if equipartition
between magnetic and total(radiation$+$gas) pressure in the disk is
assumed (i.e., $\beta=0.5$). As discussed in Section 2.2,
$B_\varphi\ll B_p$ is required for the fast moving jets from the
disk, and we adopt $\xi_\varphi=0.1$ ($B_\varphi=\xi_\varphi B_p$)
in most of our calculations. For comparison, we also plot the
results calculated with $B_\varphi=B_p$ in Figures \ref{compare} and
\ref{compareb}. We find that the maximal jet power can be extracted
with model A is insensitive to the value of $\xi_\varphi$ adopted,
because most of the gravitational energy released in the accretion
disk is transported into the jets by the magnetic field. The results
of the calculations with $\xi_\varphi=0.1$ are given in all other
plots.

The initial black hole spin parameter $a_0$ can be higher than
$0.1$, for example, a rapidly rotating black hole can be formed
after a merger of the black hole binary with mass ratio approaching
unity \citep{2003ApJ...585L.101H}. Thus, we also calculate the
problem with a relatively high initial black hole spin parameter
$a_0=0.95$ in Figure \ref{compareb}. It is not surprising that the
observed feature of the cavity can be reproduced by the calculations
with model A, as much energy can be extracted for a faster rotating
black hole. One can find that the total energy released in the jets
can be as high as $\sim 10^{62}$~erg if $a_0=0.95$ and $\tau_{\rm
Q}=10^8$~years, even for model B, i.e., the energy density of the
large-scale magnetic field is scaled with $p_{\rm tot}(H/r)^{-2}$
(see Equations \ref{lbzlivio} and \ref{lbplivio} in Section 2.2),
based on the scenario that the strength of the magnetic field
generated by dynamo processes with typical size $\lambda$ decays
with $\lambda^{-1}$ \citep{1996MNRAS.281..219T}. If this is the
case, the life timescale of this AGN $t_{\rm Q}\simeq 10\tau_{\rm
Q}=10^9$~years, which is roughly consistent with the estimates in
the previous works
\citep*[e.g.,][]{2005Natur.433..604D,h2009,2010ApJ...719.1315K,2010ApJ...725..388C}.
This implies that the duty cycle of radio activity is around $0.1$,
if model B is responsible for the jet formation in this source.
Besides the time-dependent mass accretion rate $\dot{m}(t)$, we
alternatively adopt a constant accretion rate in our calculations,
$\dot{m}=0.25$, which is the typical value of broad-line AGNs
\citep*[e.g.,][]{2006ApJ...648..128K}. We find that the results
change little if a different value of mass accretion rate in the
range of $0.1-0.5$ is adopted. The results obtained with a constant
accretion rate are similar to those carried out with the
time-dependent mass accretion rate, i.e., the jets can provide
sufficient energy to inflate the cavity of MS 0735+7421 within the
timescale of $10^8$ years with model A or model B if an initial
black hole spinning at $a_0=0.95$ (see Figure \ref{compare2}).


We compare the relative importance of the BP and BZ mechanisms in
Figure \ref{bzbp}. Similar to that suggested by \citet{l1999}, we
find that the BP power always dominates over the BZ power even if
the black hole is rotating rapidly. The reason is that the area of
the disk launching the jets is significantly larger than the surface
area of the black hole, and the strength of the field threading the
horizon of the hole is comparable with that in the inner region of
the disk \citep*[see][for the detailed discussion]{l1999}. The
initial black hole spin $a_0$ can change the power output of both
the BZ and BP mechanisms as shown in Fig. \ref{bzbp}. The jet power
produced by BZ mechanism can vary over one order of magnitude, which
is consistent with the results of MHD simulation for a thin disk in
\citet{t2010}. Black hole spin can play a more important role in the
jet formation for thicker accretion disks \citep{t2010}. The
large-scale magnetic field is dragged in by the advection dominated
accretion flow (ADAF), and \citet{c2011} found that the magnetic
field strength of the flow near the black hole horizon can be more
than one order of magnitude higher than that in the ADAF at
$\sim6GM/c^2$, due to its large radial velocity of the accretion
near the black hole horizon. It is still unclear if such effect is
important in the thin accretion disk, which is beyond the scope of
this work. We compare the results with different values of $\beta$
adopted in Figure \ref{beta}. The jet power decreases significantly
with increasing $\beta$. In most of our calculations, $\beta=0.5$ is
adopted, which is almost the upper limit on the magnetic field
strength, and then the jet power. Our main conclusions will not be
altered even if a lower $\beta$ is adopted, while the total energy
derived with model B may be insufficient for the energy stored in
the cavity of MS 0735+7421 provided a lower $\beta$ is adopted. Our
results show that strong magnetic field nearly equipartition with
the total pressure (radiation$+$gas pressure) in the accretion disk
is indeed required at least for the most energetic cavity of MS
0735$+$7421, which provides a useful clue on the formation of the
large-scale magnetic field in the accretion disk. For other less
energetic cavities, the magnetic field strength can be significantly
lower than the equipartition value, and less strict constraints are
set on the jet formation mechanisms.

It is found that the final spin parameter $a_{\rm f}$ is always very
high ($a_{\rm f}>0.9$) for all the jets that can provide sufficient
energy for the cavity of MS 0735+7421 (also see Figure
\ref{compare}), either for a low or high initial spin parameter
$a_0$ ($a_{\rm f}=0.923$ and $0.998$ for $a_0=0.1$ and $a_0=0.9$
respectively). This means that the central black hole of the AGN in
MS 0735+7421 should be rotating very rapidly now.

\section*{acknowledgements}

We thank the referee for his/her helpful comments. This work is
supported by the National Basic Research Program of China (grant
2009CB824800), the NSFC (grants 10903021, 11173043, 10821302 and
10833002), the Science and Technology Commission of Shanghai
Municipality (10XD1405000), and the CAS/SAFEA International
Partnership Program for Creative Research Teams (KJCX2-YW-T23).

\begin{figure}
\includegraphics[width=15cm]{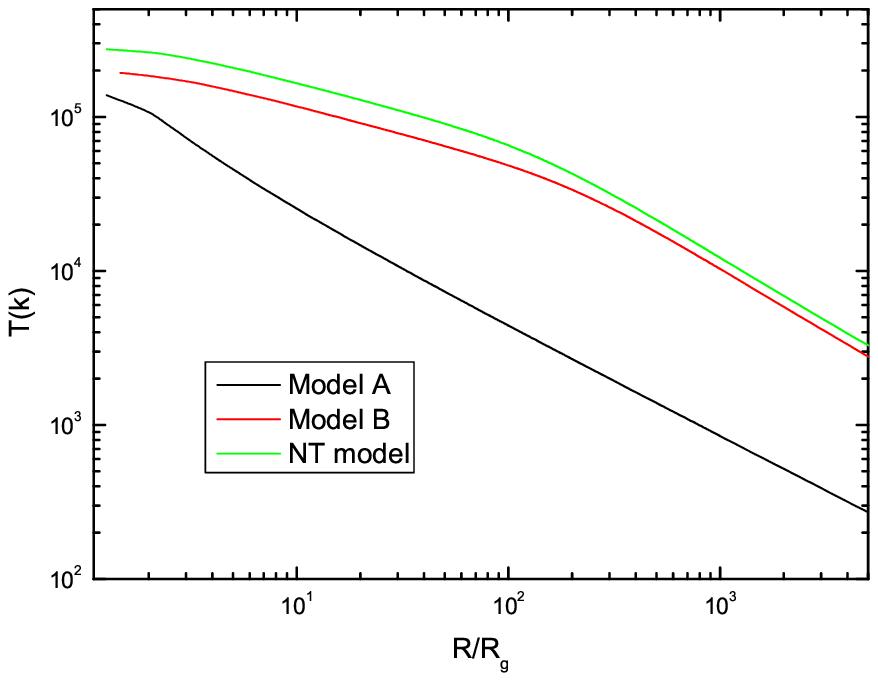}
\caption{The disk temperature of a relativistic thin accretion disk
with magnetically driven outflows, where $a_0=0.95, \beta=0.5$ and
$\dot{m}=0.25$ are adopted. The black and red lines are for the
results calculated with Model A and Model B respectively. We also
plot the results of a relativistic thin accretion disk without
outflows for comparison (green lines).  \label{property}}
\end{figure}

\begin{figure}
\includegraphics[width=15cm]{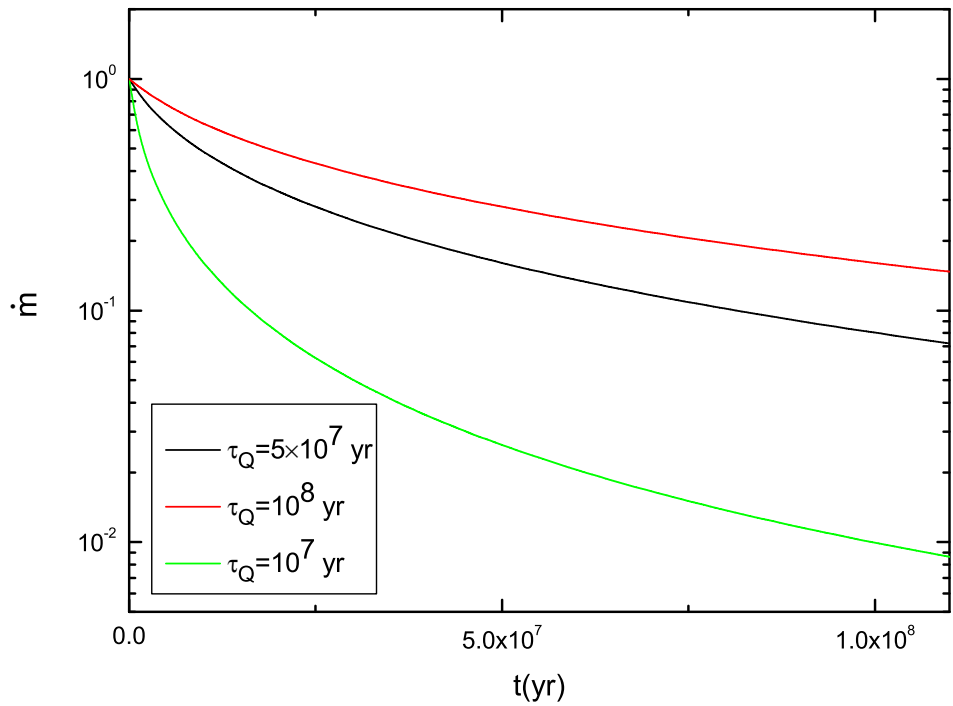}
\caption{The evolution of accretion rate as functions of outburst
time. The colored lines correspond to different values of $\tau_{\rm
Q}$, $\tau_{\rm Q}=5\times10^7$(black), $10^8$(red), and $10^7$
years(green), respectively. In all our calculations, the initial
black hole mass $M_0$ is chosen in such a way to let the final black
hole mass be $5 \times 10^9 M_{\odot}$. \label{edd}}
\end{figure}

\begin{figure}
\includegraphics[width=15cm]{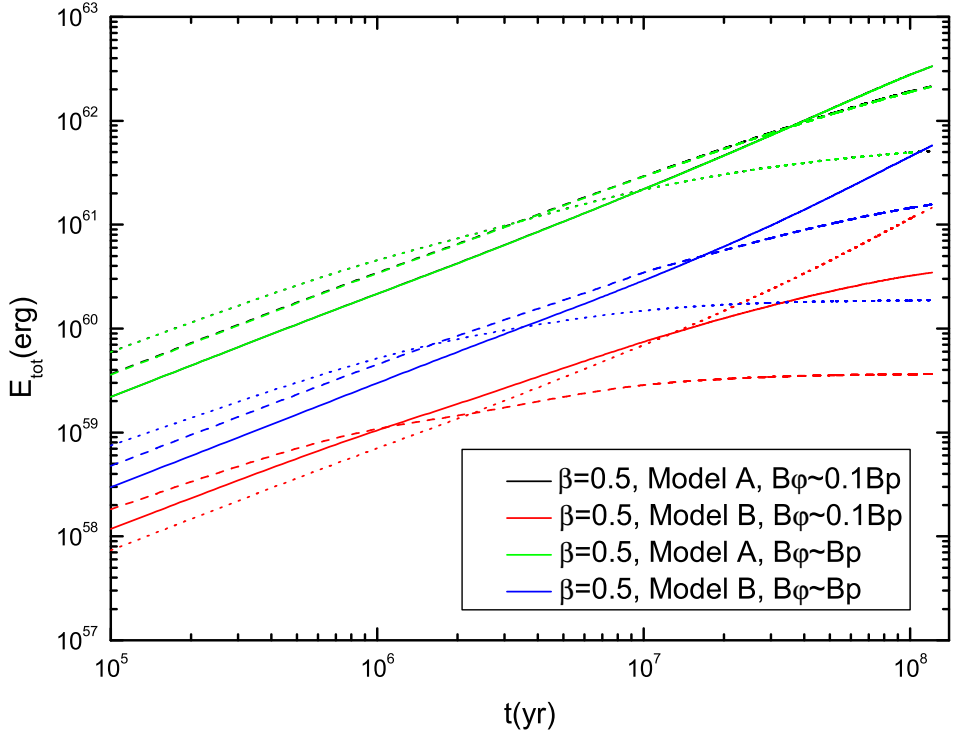}
\caption{The energy released during the outbursts as functions of
the outburst time with $a_0=0.1$ and $\beta=0.5$. The different
color lines represent the results calculated for different jet
formation models and different values of azimuthal component of the
field at the disk surface (see the Figure). The solid, dashed and
dotted line are for the results with $\tau_{\rm Q}=10^8$,
$5\times10^7$, and $10^7$ years, respectively. \label{compare}}
\end{figure}

\begin{figure}
\includegraphics[width=15cm]{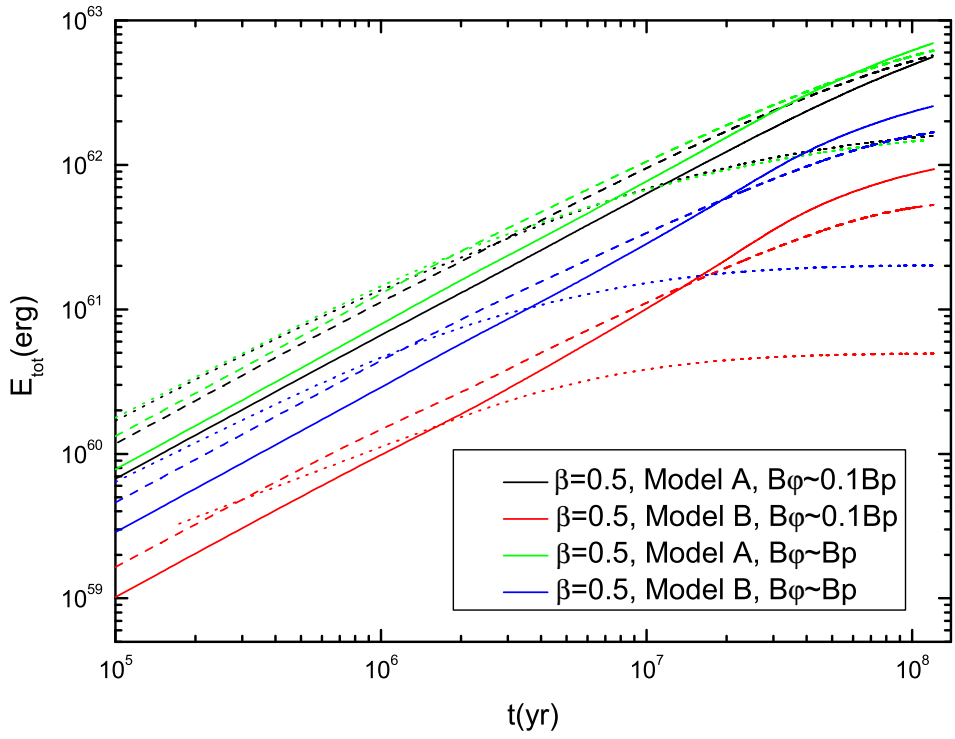}
\caption{The same as Fig. \ref{compare}, except that the initial
spin $a_0=0.95$ is adopted. \label{compareb}}
\end{figure}

\begin{figure}
\includegraphics[width=15cm]{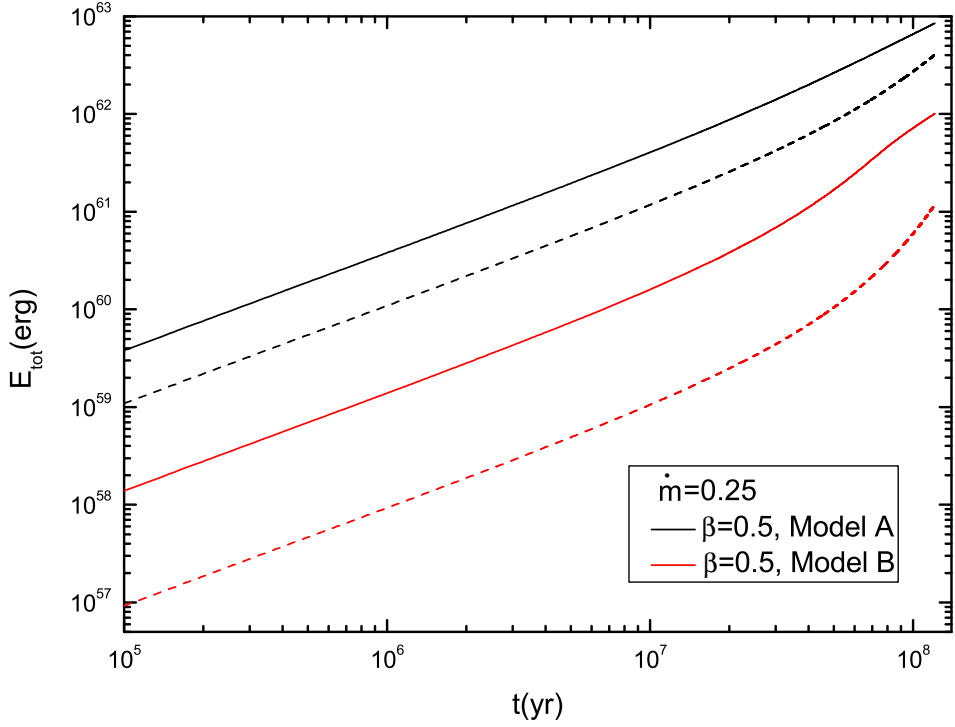}
\caption{The energy released during the outbursts as functions of
outburst time with the initial spin parameter, $a_0=0.1$(dashed) and
$0.95$(solid), respectively. The model parameter $\beta=0.5$,
$\xi_\varphi=0.1$, and a constant accretion rate $\dot{m}=0.25$ are
adopted. The colored lines are for the results calculated with
different jet formation models, Model A(black) and B(red).
\label{compare2}}
\end{figure}

\begin{figure}
\includegraphics[width=15cm]{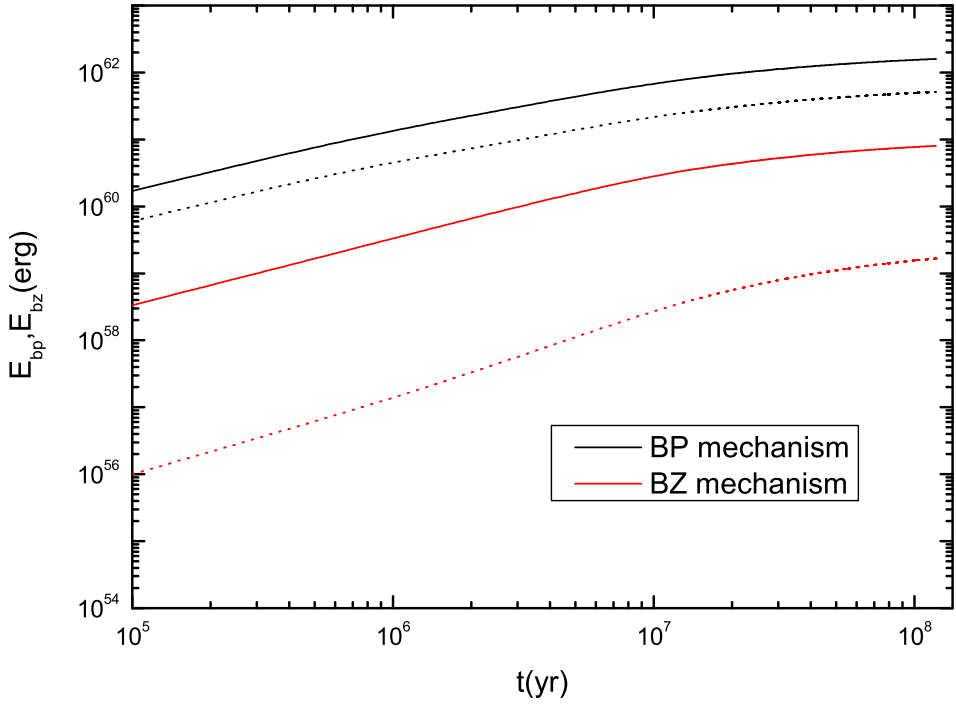}
\caption{The energy released by BZ$+$BP mechanisms (model A) as
functions of outburst time, where $\beta=0.5$, $\xi_\varphi=0.1$,
and $\tau_{\rm Q}=1\times10^7$ years are adopted. The black and red
lines are the results calculated for the BP and BZ mechanisms
respectively. The solid and dotted lines correspond to different
initial conditions, $a_0=0.95$ and $a_0=0.1$, respectively.
\label{bzbp}}

\end{figure}

\begin{figure}
\includegraphics[width=15cm]{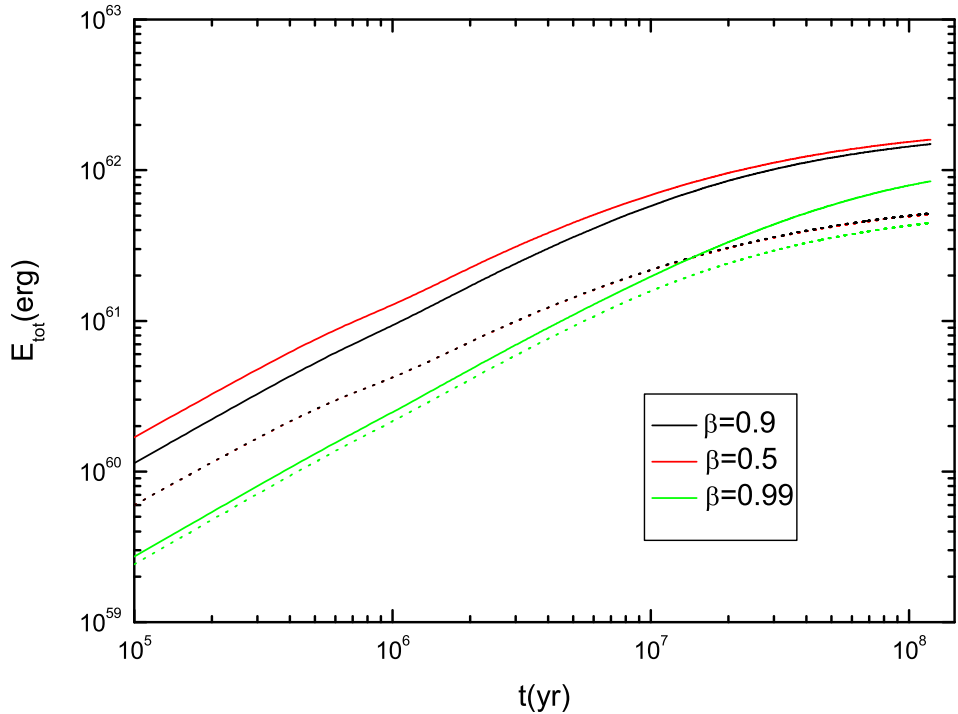}
\caption{The energy released during the outbursts as functions of
outburst time with different values of $\beta$ for model A, where
$\xi_\varphi=0.1$, $\tau_{\rm Q}=10^7$ years are adopted in the
calculations. The colored lines are for the results calculated with
$\beta=0.99$(green), $0.9$(black), and $0.5$(red), respectively. The
solid and dotted lines correspond to different initial conditions,
$a_0=0.95$ and $a_0=0.1$, respectively. \label{beta}}
\end{figure}

\begin{figure}
\includegraphics[width=15cm]{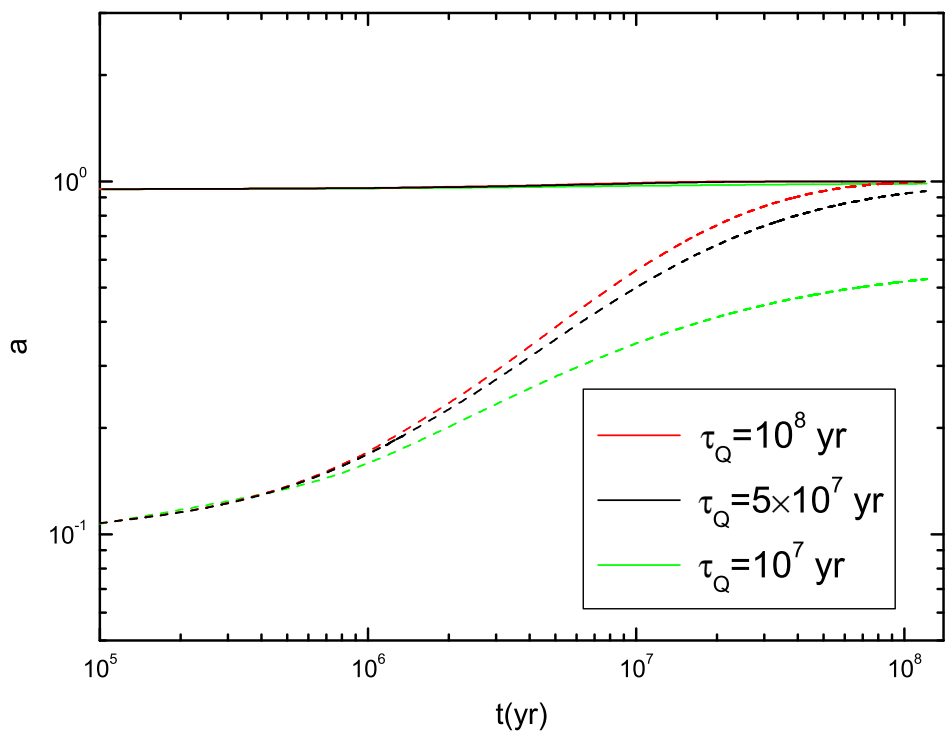}
\caption{The evolution of black hole spin as functions of outburst
time for model A, where $\beta=0.5$ and $\xi_\varphi=0.1$ are
adopted. The colored lines correspond to different values of
$\tau_{\rm Q}$, $\tau_{\rm Q}=5\times10^7$(black), $10^8$(red), and
$10^7$ years(green), respectively. The dashed and solid lines
correspond to different initial conditions, $a_0=0.1$ and
$a_0=0.95$, respectively. \label{spin2}}
\end{figure}

\end{document}